%% file: archive-recrawling-framework.tex
\documentclass{sig-alternate}

\usepackage[T1]{fontenc}
\usepackage[utf8]{inputenc}
\usepackage{booktabs}
\usepackage{graphicx}
\usepackage[colorlinks, linkcolor=black, citecolor=black, urlcolor=black]{hyperref}
\usepackage{tikz,relsize}

\begin{document}

\title{Analyzing Web archives through Topic and Event Focused Sub-Collections}

\numberofauthors{3}
\author{
  Gerhard Gossen\\
  \affaddr{L3S Research Institute}\\
  \affaddr{Leibniz Universität Hannover}\\
  \affaddr{Appelstraße 9a}\\
  \affaddr{30167 Hannover, Germany}\\
  \email{ gossen@L3S.de}
  \alignauthor
  Elena Demidova \\
  \affaddr{Web and Internet Science Group, Electronics \& Computer Science}\\
  \affaddr{University of Southampton} \\
  \email{e.demidova@soton.ac.uk}
  \alignauthor
  Thomas Risse\\
  \affaddr{L3S Research Institute}\\
  \affaddr{Leibniz Universität Hannover}\\
  \affaddr{Appelstraße 9a}\\
  \affaddr{30167 Hannover, Germany}\\
  \email{risse@L3S.de}
}

\CopyrightYear{2016}
\setcopyright{acmlicensed}
\conferenceinfo{WebSci '16,}{May 22 - 25, 2016, Hannover, Germany}
\isbn{978-1-4503-4208-7/16/05}\acmPrice{\$15.00}
\doi{http://dx.doi.org/10.1145/2908131.2908175}

\maketitle

\begin{abstract}
Web archives capture the history of the Web and are therefore an important source to study how societal developments have been reflected on the Web.
However, the large size of Web archives and their temporal nature pose many challenges to researchers interested in working with these collections.
In this work, we describe the challenges of working with Web archives and propose the research methodology of extracting and studying sub-collections of the archive focused on specific topics and events.
We discuss the opportunities and challenges of this approach and suggest a framework for creating sub-collections.
\end{abstract}




\input{tab-scopes}

\section{Introduction}
\label{sec:introduction}

Web archives such as the Internet Archive\footnote{\url{http://www.archive.org}} or the archives collected by national libraries
allow researchers in Web Science and the Digital Humanities to look back at the past of the Web and trace its development over time.
These archives are created by regularly crawling the Web (in the case of the Internet Archive) or selected subsets (typically national sub-domains) to create snapshots of Web sites at different points in time.
Researchers can look up any of the crawled versions to look back at specific points in the past or compare different versions.

An important challenge when using Web archives is the access to the collected data.
As an example, the Internet Archive has a size of several petabytes over a time span of 20 years.
A researcher working with such an archive needs efficient and effective tools to scope relevant documents for further research.
In contrast to typical use cases where only individual pages are considered (e.g. in legal disputes or to provide persistent links) or the entire archive is analyzed using automatic methods (e.g. using text mining) \cite{reynolds2013},
many research questions in Web Science and the Digital Humanities require an analysis of documents related to specific topics and events.
The analysis is often performed manually, therefore the documents to be analyzed in detail have to be carefully selected.
We call a collection of documents related to a specific topic or event a \textbf{topic} and \textbf{event focused sub-collection} of the archive.

Current tools do not support the researcher enough in creating such a sub-collection (cf. for example~\cite{deswarte2015}).
Current approaches use \emph{browsing} and \emph{searching} as access methods for Web archives.
\emph{Browsing} is done by entering URLs in a Web interface such as the Wayback Machine~\cite{wayback} and navigating the archived pages using hyperlinks.
This requires that the URLs of relevant pages or at least pages linking to them are known in advance.
As the entire process is done manually, the researcher will typically have to try many hyperlinks that are not available in the archive.
The researcher also has to be aware of temporal drift while navigating the archive \cite{ainsworth2013}.
Temporal drift occurs because the linking and linked page were usually crawled at different times and therefore each navigation step moves the analyzed point in time.
Iterated navigation can even lead to page versions that were crawled outside of the relevant time window.
The browsing approach is therefore inefficient and error-prone.

An alternative approach is to use keyword \emph{search} over the entire archive.
Here the user only needs to enter keywords into a search interface and can see all pages matching the query.
Many search interfaces also provide faceted browsing, where the search can be narrowed down further by e.g. the crawl time, the document content type or the domain name.
It is much easier for users to get started using this approach, as they do not have to know about relevant documents in advance.
They can also combine this approach with the browsing approach by starting the navigation from a search result document.
Search has however many technical challenges.
First of all, the archive needs to create and keep up to date a full text index of its entire content and execute queries over this index.
Because of the typically large size of the archives this necessitates distributed indexing and retrieval architectures \cite{costa2013} that require many computational resources as well as a lot of technical expertise to set up and maintain.
An additional issue is the ranking of documents matching a query.
On the one hand, the archive contains many snapshots of the same document, which can negatively impact traditional ranking methods that use for example link-based measures \cite{Costa:2014:LTR:2600428.2609619}.
On the other hand, the information need of Web archive users is different than in standard search applications because it is usually focused around specific time periods and therefore requires different ranking measures \cite{weikum2011}.
Therefore the searching of Web archive using current tools still requires a lot of manual effort from the user to tune queries and go through result lists.

We therefore propose an alternative approach for the access to Web archives through the automatic extraction of topic and event focused sub-collections.
Such sub-collections contain documents from the archive that have been automatically classified as referring to a given topic or event.
We may additionally pose the constraints that documents in the collection are connected through hyperlinks or that the temporal distance between any two documents in the collection is minimal.
By extracting such sub-collections we provide researchers with relevant sets of documents, which can be further analyzed in their appropriate context.

In this work we define the concept of topic and event focused sub-collections, describe a framework for extracting them and discuss challenges.


\section{Web Archive Sub-Collections}
\label{sec:web-archive-sub}

In this section we define the concept of topic and event focused sub-collections and describe several important variants of such sub-collections.

A \textbf{Web archive} is a collection of \textbf{Web document snapshots}.
Web document snapshot refers to the content retrieved from a given URL (\textbf{document URL}) at a given time (\textbf{crawl time}).
In addition to the content, the archive typically also stores metadata about the snapshot such as the software used for retrieval, the HTTP headers or the document content type.

An \textbf{topic and event focused sub-collection} is a set of Web document snapshots, where each snapshot is available in the given Web archive.
It is defined in a \textbf{sub-collection specification} that describes the scope of the sub-collection.
The format of the sub-collection specification depends on the \textbf{extraction algorithm} used to create the sub-collection.
Each type of algorithm defines a number of \textbf{scopes} that describe relevant documents.
A list of exemplary scopes is given in Tab.~\ref{tab:scopes}.
Scopes can be combined to narrow down the sub-collection.
For example, a simple algorithm that supports the \texttt{URL} and \texttt{time} scopes can be used to extract all snapshots of a given URL in a specific time frame.
A scope does not need to be exact.
For example, a topic scope can be implemented using a machine learning algorithm, that classifies a snapshot as relevant based e.g. on the similarity to a given set of topic keywords.
In this case evaluation metrics like precision and recall can be used to analyze the quality of a given algorithm.

Given the large size of typical Web archives it is often neither feasible nor desired to find all snapshots matching a scope.
Therefore an algorithm should have a high precision in matching the scopes but may have a lower recall.
A good algorithm should however aim to find a representative sub-collection, i.e. one that has a similar diversity as the original archives in terms of domains, crawl times or types of sources.

We further distinguish between a \textbf{connected} and a \textbf{disconnected} sub-collection.
A connected sub-collection needs to contain for any snapshot $s$ contained in the sub-collection also at least one snapshot $t$ for each document that is linked from $s$, if one is available in the archive.
In contrast, a disconnected sub-collection can consist only of isolated snapshots.
A connected sub-collection is needed to perform e.g. link graph analyses, whereas e.g. content-based analyses can also be performed on a disconnected sub-collection.

An additional distinction is between \textbf{snapshot} and \textbf{timeline} sub-collections.
In a snapshot sub-collection, each document URL should occur only once,
a timeline sub-collection should however have all snapshots of an in scope URL that are also in scope.
A snapshot sub-collection is useful in synchronic analyses, where the researcher is looking at a specific point in time and does not want to deal with multiple versions of the same URL.
In contrast, a timeline sub-collection is needed to perform diachronic analyses where we want to track a development over time.

\section{Sub-Collection Extraction\\ Framework}
\label{sec:framework}

  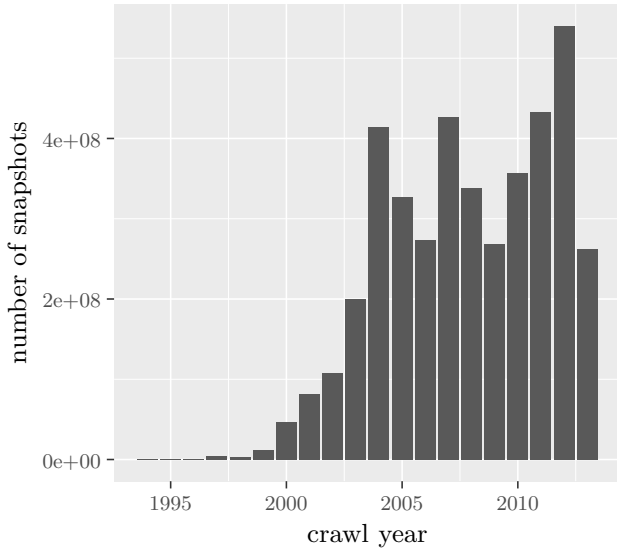
\begin{figure}[t]
    \centering \input{figs/counts}
    \caption{Number of snapshots by crawl time}
    \label{fig:count}
  \end{figure}

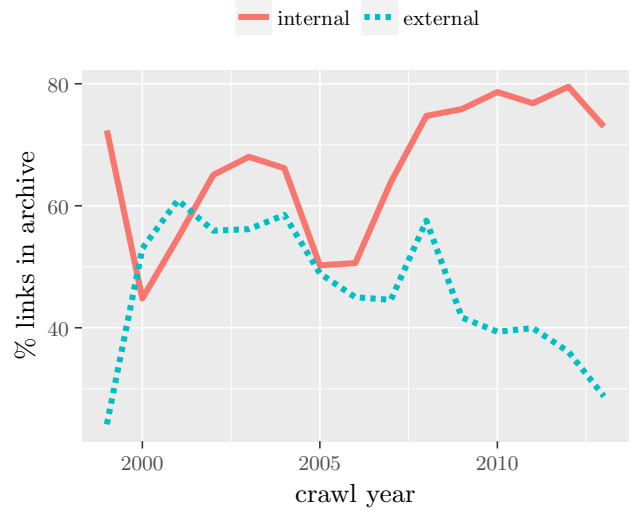
\begin{figure}[t]
  \centering \input{figs/ratios}
  \caption{Average rate of outlinks that are contained in the archive.
    Internal links have the same domain name as the linking page,
    external links go to different domains.}
  \label{fig:ratios}
\end{figure}
In this section we will describe the framework, in which the sub-collection extraction process takes place and describe measures to evaluate extraction algorithms.

To create a Web archive sub-collection $C$ the researcher first has to choose a base Web archive $W$ and create a sub-collection specification $CS$ that describe their collection need.
Then they need to select an algorithm $A$ that supports the scopes specified in $CS$ and run it over the archive $W$, using the sub-collection specification $CS$ as a parameter.
The result of this process is the extracted sub-collection $C$.
We expect that the extraction process will typically be iterative, i.e. that the research will create a modified specification $CS'$ after analyzing the sub-collection $C$ and create a new sub-collection $C'$, maybe even using a different extraction algorithm $A'$.

To compare different extraction algorithms, we can use the following evaluation measures:
\begin{description}
\item[Precision]
As there is no metadata about what topics and events a snapshot in a Web archive is relevant for, the relevance calculation needs to be done using automated methods such as machine learning, e.g. Support Vector Machines (SVMs) for topic classification~\cite{Joachims1998}.
These methods will however mistake some irrelevant documents as relevant and vice versa.
The precision measures the rate of such errors (cf.~\cite{manning2008}):
\[\text{precision} := \frac{|\text{retrieved relevant snapshots}|}{|\text{retrieved snapshots}|} \]
\item[Recall]
An extraction algorithm can simply iterate over the entire archive and evaluate the specified scopes on each snapshot.
Given the large size of Web archives, this is often prohibitively expensive even if parallel processing facilities are available.
Therefore it is desirable that the extraction algorithm can use indexes or heuristics to speed up the execution.
The recall measures the fraction of relevant snapshots that were extracted from the archive (cf.~\cite{manning2008}):
\[\text{recall} := \frac{|\text{retrieved relevant snapshots}|}{|\text{relevant snapshots}|} \]
In this way it quantifies the expected loss of using a more efficient algorithm.

Note that we also need to consider the precision of the relevance estimation method when computing the recall as it may perform differently on the selected subset, for example because the extraction algorithm will preferentially select pages from certain domains or having a certain link structure.
In this case also the recall on the entire collection needs to be examined.
\item[Diversity] 
As described above, the goal of extracting sub-collections is to find a set of documents that help in answering a research question.
While the collection needs to have a manageable size so that it can be analyzed, it also needs to be representative of the entire archive.
This can be measured using diversity measures that describe how well different aspects of the given topic or event are represented \cite{Clarke:2008:NDI:1390334.1390446}.
\item[Link completeness] 
When we analyze the context of a snapshot, e.g. using link graph analysis, it is important to also have all relevant linked pages in the sub-collection.
We measure the link completeness of a collection $C$ as follows:
\[ lc(C) = \mathlarger{\sum_{s \in C}}\frac{|\text{retrieved relevant outlinks of $s$}|}{|\text{relevant outlinks of $s$}|} \]

\item[Temporal coherence] 
Snapshots in a Web archive are typically crawled at different points in time, even if they refer to the same event.
Additionally, a given URL may have been crawled several times in a relevant time frame, providing an algorithm with multiple snapshots to choose from.
The selection of snapshots can however introduce errors in the downstream analyses when selecting snapshots that are from distant points in time.
A way to reduce this risk is to optimize the selection of snapshot such that the time between any pair of snapshots is minimized.
A similar measure is the \emph{blur} of a Web archive which also considers the expected number of changes to retrieved pages \cite{denev2011}.

\item[Run time]
To allow for a fast iteration of refined sub-collection specifications, it is important that extraction algorithms can produce their results fast.
As operations on Web archives are often executed in parallel on large clusters, typical run time measures such as the elapsed time in seconds are less useful because they are heavily influenced by the size and current load of the cluster.
A better evaluation metric is instead the number of disk accesses to retrieve and evaluate snapshots, as this is typically the dominant cost in the extraction process.
\end{description}

\begin{figure}[t]
  \centering \input{figs/features}
  \caption{Average number occurrences of HTML tags per document. Each
    value is normalized to its maximum value.}
  \label{fig:tags.n}
\end{figure}
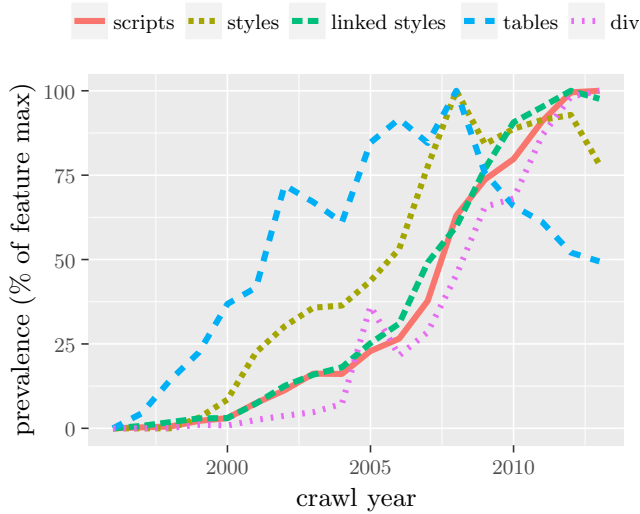

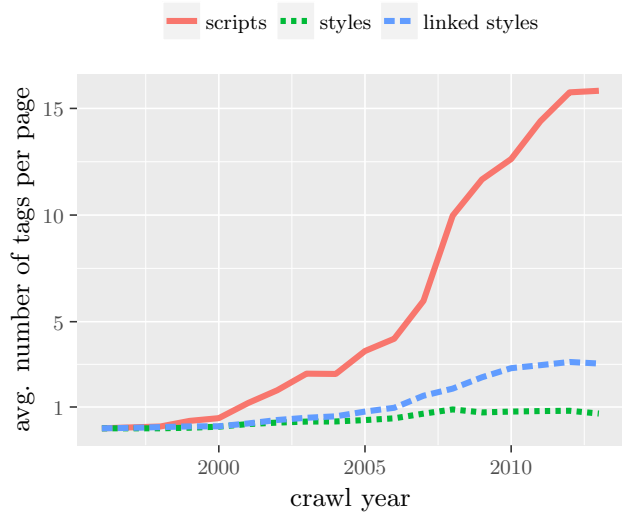
\begin{figure}[t]
  \centering \input{figs/features_t}
  \caption{Average number occurrences of HTML tags per document.}
  \label{fig:tags}
\end{figure}

\begin{figure}[t]
  \centering \input{figs/links}
  \caption{Average number of outlinks per document. Internal links
    have the same domain name as the linking page, external links go
    to different domains.}
  \label{fig:links}
\end{figure}
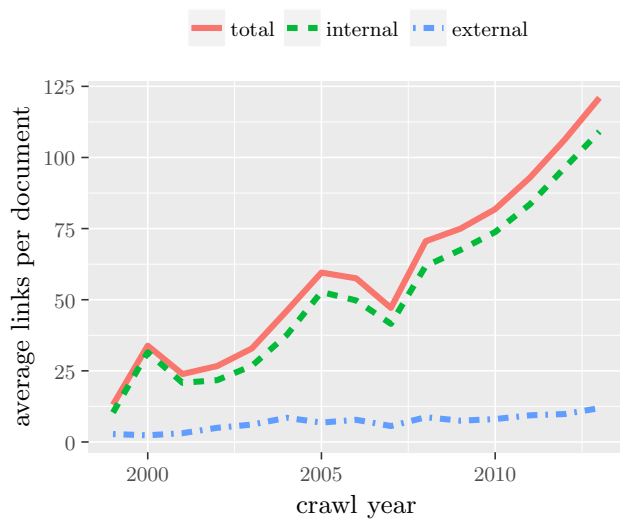

\section{Challenges}
\label{sec:challenges}
In this section we discuss several challenges when working with Web archives, especially when extracting sub-collections.
We illustrate the challenges using data extracted from the German Web archive, a collection of all Web pages from the \texttt{.de} top level domain crawled by the Internet Archive between 1994 and 2013.
All values except the total number of snapshots (Fig.~\ref{fig:count}) were calculated using a random sample of 40K snapshots.

\begin{description}
\item[Temporal Scope]
  The number of snapshots per year can have strong fluctuations (see Fig.~\ref{fig:count}).
  In general there are more snapshots for recent years, which is consistent with the general trend of a growing web.
  The exact number of snapshots per year can however vary due to different crawl strategies, intermittent errors and other factors.
%

  In the context of Web archive sub-collections this means that a temporal scope can only be used effectively for some time periods, whereas e.g. in the first years of the archive's time span it may exclude too many documents.
  This also means that diversification may be necessary to avoid that snapshots from sparse time periods get lost while more active time periods are over-represented.
\item[Archive Completeness]
  Most archives do not have complete snapshots of the Web due to limited resources, legal restrictions and technical challenges \cite{Kelly:2014:AAT:2740769.2740774} that restrict the collection of the archives.
  Fig.~\ref{fig:ratios} gives estimates of rate of missing content.
  For each snapshot in our sample we extracted the outgoing links and tried to retrieve them from the archive.
  We see that about 50-80\% of all links within the same domain (internal links) can be retrieved, whereas for external links this rate is much lower at 20-60\%.
  A possible reason for the lower ratio for external links is the restriction of the archive to the \texttt{.de} domain.
  This is however a realistic scenario, as many current Web archives are run by national libraries and have similar restrictions to Web sites from specific countries or top level domains.

  For the sub-collection extraction this means that there is an inherent upper bound on the achievable link completeness.
  It also suggests that optimizing for link completeness may bias the sub-collection towards sites with many relevant internal links and away from hub pages with many relevant external links, as the former are likely to be more complete.
\item[Content Diversity]
  The content in our archive spans 20 years, which means that it reflects many developments of Web technology.
  Fig.~\ref{fig:tags.n} shows the relative frequency of representative HTML tags over time.
  Each curve has been normalized such that it reflects the prevalence of the tag in comparison to its peak value.
  We can easily see the decline of table-based layouts in the 2000s and the growth of layout techniques using \texttt{div} tags styled by CSS style sheets.
  Similarly, Fig.~\ref{fig:tags} shows the absolute average number of scripts and style sheets per page.
  We see after 2005 a dramatic increase in the number scripts and to a lesser extent of linked style sheets.
  The former may be a reflection of the increased use of advertising networks and tracking services.

  Similarly, we can look at the number of links per page over time (Fig.~\ref{fig:links}).
  Whereas the number of external links stays relatively stable, the number of internal links increases continuously.
  This may be due to the maturing of Web sites that accumulate content over time, but it may also reflect the changed Web environment where e.g. search engine optimization (SEO) through specific link strategies becomes more common.

  For the extraction of sub-collections this means that our algorithms must be adaptive to different types of Web content:
  Over time the format of Web pages has changed dramatically, therefore also the position of relevance cues on the pages may have changed.
  Furthermore, the value of links has changed over time, such that we have to be more selective when selecting links on more recent pages.
\end{description}

\section{Conclusion}
\label{sec:conclusion}
In this work we have presented a new approach for the access to Web archives through the automatic extraction of topic and event focused sub-collections.
In contrast to existing approaches, sub-collection decrease the amount of manual effort required to find a reasonably-sized collection of documents for further research, while increasing the value of the collection through better extraction of link graphs and the avoidance of temporal drift.
We have defined the problem of extracting sub-collections and have described several typical extensions to the problem.
Additionally, we have described the framework in which this approach is applied and have shown several evaluation metrics to compare algorithms for this problem.
Based on data from a real-world Web archive we have demonstrated several issues for algo-\\\newpage\noindent rithms trying to solve this problem and have discussed approaches for dealing with them.
We hope that this work sparks interest in the extraction and use of topic and event focused sub-collections.
In future work, we will present algorithms that create sub-collections following this framework.

\section*{Acknowledgments}
\small{
 This work was partially funded by the European Research Council under ALEXANDRIA (ERC 339233) and the European Commission under SoBigData (RIA 654024) and H2020-MSCA-ITN-2014 WDAqua (grant agreement 64279).}

\bibliographystyle{abbrv}
\bibliography{bibliography}

\end{document}

%% file: tab-scopes.tex
\begin{table*}
  \centering
  \begin{tabular}{llp{10cm}}
    \toprule
    Scope        &          Type  & Description \\
    \midrule
    URL          & list of URLs   & documents that need to be in the sub-collection \\ 
    domain       & list of domains & domains that the sub-collection should be restricted to \\
    time         & time interval  & relevant timeframe \\
    keywords     & list of keywords & descriptive keywords for the sub-collection topic \\
    event/entity & list of knowledgebase references & entries in a knowledgebase such as FreeBase \cite{bollacker2008freebase} that are the topic of the sub-collection\\
    size         & number of documents & target size of the sub-collection \\
    \bottomrule
  \end{tabular}
  \caption{Exemplary scopes used in a sub-collection specification. This list is not exhaustive.}
  \label{tab:scopes}
\end{table*}

%% file: figs/counts.tex
\begin{tikzpicture}[x=1pt,y=1pt]
\definecolor{fillColor}{RGB}{255,255,255}
\path[use as bounding box,fill=fillColor,fill opacity=0.00] (0,0) rectangle (245.72,216.81);
\begin{scope}
\path[clip] (  0.00,  0.00) rectangle (245.72,216.81);
\definecolor{drawColor}{RGB}{255,255,255}
\definecolor{fillColor}{RGB}{255,255,255}

\path[draw=drawColor,line width= 0.6pt,line join=round,line cap=round,fill=fillColor] (  0.00,  0.00) rectangle (245.72,216.81);
\end{scope}
\begin{scope}
\path[clip] ( 48.58, 30.69) rectangle (240.22,211.31);
\definecolor{fillColor}{gray}{0.92}

\path[fill=fillColor] ( 48.58, 30.69) rectangle (240.22,211.31);
\definecolor{drawColor}{RGB}{255,255,255}

\path[draw=drawColor,line width= 0.3pt,line join=round] ( 48.58, 69.28) --
	(240.22, 69.28);

\path[draw=drawColor,line width= 0.3pt,line join=round] ( 48.58,130.04) --
	(240.22,130.04);

\path[draw=drawColor,line width= 0.3pt,line join=round] ( 48.58,190.81) --
	(240.22,190.81);

\path[draw=drawColor,line width= 0.3pt,line join=round] ( 91.87, 30.69) --
	( 91.87,211.31);

\path[draw=drawColor,line width= 0.3pt,line join=round] (135.64, 30.69) --
	(135.64,211.31);

\path[draw=drawColor,line width= 0.3pt,line join=round] (179.42, 30.69) --
	(179.42,211.31);

\path[draw=drawColor,line width= 0.3pt,line join=round] (223.19, 30.69) --
	(223.19,211.31);

\path[draw=drawColor,line width= 0.6pt,line join=round] ( 48.58, 38.90) --
	(240.22, 38.90);

\path[draw=drawColor,line width= 0.6pt,line join=round] ( 48.58, 99.66) --
	(240.22, 99.66);

\path[draw=drawColor,line width= 0.6pt,line join=round] ( 48.58,160.43) --
	(240.22,160.43);

\path[draw=drawColor,line width= 0.6pt,line join=round] ( 69.98, 30.69) --
	( 69.98,211.31);

\path[draw=drawColor,line width= 0.6pt,line join=round] (113.76, 30.69) --
	(113.76,211.31);

\path[draw=drawColor,line width= 0.6pt,line join=round] (157.53, 30.69) --
	(157.53,211.31);

\path[draw=drawColor,line width= 0.6pt,line join=round] (201.30, 30.69) --
	(201.30,211.31);
\definecolor{fillColor}{gray}{0.35}

\path[fill=fillColor] ( 57.29, 38.90) rectangle ( 65.17, 38.90);

\path[fill=fillColor] ( 66.04, 38.90) rectangle ( 73.92, 38.90);

\path[fill=fillColor] ( 74.80, 38.90) rectangle ( 82.68, 39.00);

\path[fill=fillColor] ( 83.55, 38.90) rectangle ( 91.43, 40.33);

\path[fill=fillColor] ( 92.31, 38.90) rectangle (100.19, 39.89);

\path[fill=fillColor] (101.06, 38.90) rectangle (108.94, 42.45);

\path[fill=fillColor] (109.82, 38.90) rectangle (117.69, 53.19);

\path[fill=fillColor] (118.57, 38.90) rectangle (126.45, 63.84);

\path[fill=fillColor] (127.33, 38.90) rectangle (135.20, 71.78);

\path[fill=fillColor] (136.08, 38.90) rectangle (143.96, 99.69);

\path[fill=fillColor] (144.83, 38.90) rectangle (152.71,164.83);

\path[fill=fillColor] (153.59, 38.90) rectangle (161.47,138.23);

\path[fill=fillColor] (162.34, 38.90) rectangle (170.22,121.95);

\path[fill=fillColor] (171.10, 38.90) rectangle (178.98,168.53);

\path[fill=fillColor] (179.85, 38.90) rectangle (187.73,141.81);

\path[fill=fillColor] (188.61, 38.90) rectangle (196.49,120.29);

\path[fill=fillColor] (197.36, 38.90) rectangle (205.24,147.47);

\path[fill=fillColor] (206.12, 38.90) rectangle (214.00,170.54);

\path[fill=fillColor] (214.87, 38.90) rectangle (222.75,203.10);

\path[fill=fillColor] (223.63, 38.90) rectangle (231.51,118.73);
\end{scope}
\begin{scope}
\path[clip] (  0.00,  0.00) rectangle (245.72,216.81);
\definecolor{drawColor}{gray}{0.30}

\node[text=drawColor,anchor=base east,inner sep=0pt, outer sep=0pt, scale=  0.88] at ( 43.63, 35.87) {0e+00};

\node[text=drawColor,anchor=base east,inner sep=0pt, outer sep=0pt, scale=  0.88] at ( 43.63, 96.63) {2e+08};

\node[text=drawColor,anchor=base east,inner sep=0pt, outer sep=0pt, scale=  0.88] at ( 43.63,157.39) {4e+08};
\end{scope}
\begin{scope}
\path[clip] (  0.00,  0.00) rectangle (245.72,216.81);
\definecolor{drawColor}{gray}{0.20}

\path[draw=drawColor,line width= 0.6pt,line join=round] ( 45.83, 38.90) --
	( 48.58, 38.90);

\path[draw=drawColor,line width= 0.6pt,line join=round] ( 45.83, 99.66) --
	( 48.58, 99.66);

\path[draw=drawColor,line width= 0.6pt,line join=round] ( 45.83,160.43) --
	( 48.58,160.43);
\end{scope}
\begin{scope}
\path[clip] (  0.00,  0.00) rectangle (245.72,216.81);
\definecolor{drawColor}{gray}{0.20}

\path[draw=drawColor,line width= 0.6pt,line join=round] ( 69.98, 27.94) --
	( 69.98, 30.69);

\path[draw=drawColor,line width= 0.6pt,line join=round] (113.76, 27.94) --
	(113.76, 30.69);

\path[draw=drawColor,line width= 0.6pt,line join=round] (157.53, 27.94) --
	(157.53, 30.69);

\path[draw=drawColor,line width= 0.6pt,line join=round] (201.30, 27.94) --
	(201.30, 30.69);
\end{scope}
\begin{scope}
\path[clip] (  0.00,  0.00) rectangle (245.72,216.81);
\definecolor{drawColor}{gray}{0.30}

\node[text=drawColor,anchor=base,inner sep=0pt, outer sep=0pt, scale=  0.88] at ( 69.98, 19.68) {1995};

\node[text=drawColor,anchor=base,inner sep=0pt, outer sep=0pt, scale=  0.88] at (113.76, 19.68) {2000};

\node[text=drawColor,anchor=base,inner sep=0pt, outer sep=0pt, scale=  0.88] at (157.53, 19.68) {2005};

\node[text=drawColor,anchor=base,inner sep=0pt, outer sep=0pt, scale=  0.88] at (201.30, 19.68) {2010};
\end{scope}
\begin{scope}
\path[clip] (  0.00,  0.00) rectangle (245.72,216.81);
\definecolor{drawColor}{RGB}{0,0,0}

\node[text=drawColor,anchor=base,inner sep=0pt, outer sep=0pt, scale=  1.10] at (144.40,  7.70) {crawl year};
\end{scope}
\begin{scope}
\path[clip] (  0.00,  0.00) rectangle (245.72,216.81);
\definecolor{drawColor}{RGB}{0,0,0}

\node[text=drawColor,rotate= 90.00,anchor=base,inner sep=0pt, outer sep=0pt, scale=  1.10] at ( 15.28,121.00) {number of snapshots};
\end{scope}
\end{tikzpicture}

%% file: figs/ratios.tex
\begin{tikzpicture}[x=1pt,y=1pt]
\definecolor{fillColor}{RGB}{255,255,255}
\path[use as bounding box,fill=fillColor,fill opacity=0.00] (0,0) rectangle (245.72,216.81);
\begin{scope}
\path[clip] (  0.00,  0.00) rectangle (245.72,216.81);
\definecolor{drawColor}{RGB}{255,255,255}
\definecolor{fillColor}{RGB}{255,255,255}

\path[draw=drawColor,line width= 0.6pt,line join=round,line cap=round,fill=fillColor] (  0.00,  0.00) rectangle (245.72,216.81);
\end{scope}
\begin{scope}
\path[clip] ( 33.42, 30.69) rectangle (240.22,171.25);
\definecolor{fillColor}{gray}{0.92}

\path[fill=fillColor] ( 33.42, 30.69) rectangle (240.22,171.25);
\definecolor{drawColor}{RGB}{255,255,255}

\path[draw=drawColor,line width= 0.3pt,line join=round] ( 33.42, 50.53) --
	(240.22, 50.53);

\path[draw=drawColor,line width= 0.3pt,line join=round] ( 33.42, 96.72) --
	(240.22, 96.72);

\path[draw=drawColor,line width= 0.3pt,line join=round] ( 33.42,142.91) --
	(240.22,142.91);

\path[draw=drawColor,line width= 0.3pt,line join=round] ( 89.82, 30.69) --
	( 89.82,171.25);

\path[draw=drawColor,line width= 0.3pt,line join=round] (156.96, 30.69) --
	(156.96,171.25);

\path[draw=drawColor,line width= 0.3pt,line join=round] (224.10, 30.69) --
	(224.10,171.25);

\path[draw=drawColor,line width= 0.6pt,line join=round] ( 33.42, 73.63) --
	(240.22, 73.63);

\path[draw=drawColor,line width= 0.6pt,line join=round] ( 33.42,119.81) --
	(240.22,119.81);

\path[draw=drawColor,line width= 0.6pt,line join=round] ( 33.42,166.00) --
	(240.22,166.00);

\path[draw=drawColor,line width= 0.6pt,line join=round] ( 56.25, 30.69) --
	( 56.25,171.25);

\path[draw=drawColor,line width= 0.6pt,line join=round] (123.39, 30.69) --
	(123.39,171.25);

\path[draw=drawColor,line width= 0.6pt,line join=round] (190.53, 30.69) --
	(190.53,171.25);
\definecolor{drawColor}{RGB}{248,118,109}

\path[draw=drawColor,line width= 2.3pt,line join=round] ( 42.82,148.35) --
	( 56.25, 84.71) --
	( 69.68,107.62) --
	( 83.11,131.52) --
	( 96.54,138.34) --
	(109.96,134.04) --
	(123.39, 97.25) --
	(136.82, 98.12) --
	(150.25,128.77) --
	(163.68,153.83) --
	(177.11,156.41) --
	(190.53,162.85) --
	(203.96,158.62) --
	(217.39,164.86) --
	(230.82,149.80);
\definecolor{drawColor}{RGB}{0,191,196}

\path[draw=drawColor,line width= 2.3pt,dash pattern=on 2pt off 2pt ,line join=round] ( 42.82, 37.08) --
	( 56.25,103.67) --
	( 69.68,121.94) --
	( 83.11,110.37) --
	( 96.54,110.97) --
	(109.96,116.34) --
	(123.39, 94.21) --
	(136.82, 85.17) --
	(150.25, 84.21) --
	(163.68,114.16) --
	(177.11, 77.50) --
	(190.53, 72.13) --
	(203.96, 73.48) --
	(217.39, 64.57) --
	(230.82, 47.68);
\end{scope}
\begin{scope}
\path[clip] (  0.00,  0.00) rectangle (245.72,216.81);
\definecolor{drawColor}{gray}{0.30}

\node[text=drawColor,anchor=base east,inner sep=0pt, outer sep=0pt, scale=  0.88] at ( 28.47, 70.60) {40};

\node[text=drawColor,anchor=base east,inner sep=0pt, outer sep=0pt, scale=  0.88] at ( 28.47,116.78) {60};

\node[text=drawColor,anchor=base east,inner sep=0pt, outer sep=0pt, scale=  0.88] at ( 28.47,162.97) {80};
\end{scope}
\begin{scope}
\path[clip] (  0.00,  0.00) rectangle (245.72,216.81);
\definecolor{drawColor}{gray}{0.20}

\path[draw=drawColor,line width= 0.6pt,line join=round] ( 30.67, 73.63) --
	( 33.42, 73.63);

\path[draw=drawColor,line width= 0.6pt,line join=round] ( 30.67,119.81) --
	( 33.42,119.81);

\path[draw=drawColor,line width= 0.6pt,line join=round] ( 30.67,166.00) --
	( 33.42,166.00);
\end{scope}
\begin{scope}
\path[clip] (  0.00,  0.00) rectangle (245.72,216.81);
\definecolor{drawColor}{gray}{0.20}

\path[draw=drawColor,line width= 0.6pt,line join=round] ( 56.25, 27.94) --
	( 56.25, 30.69);

\path[draw=drawColor,line width= 0.6pt,line join=round] (123.39, 27.94) --
	(123.39, 30.69);

\path[draw=drawColor,line width= 0.6pt,line join=round] (190.53, 27.94) --
	(190.53, 30.69);
\end{scope}
\begin{scope}
\path[clip] (  0.00,  0.00) rectangle (245.72,216.81);
\definecolor{drawColor}{gray}{0.30}

\node[text=drawColor,anchor=base,inner sep=0pt, outer sep=0pt, scale=  0.88] at ( 56.25, 19.68) {2000};

\node[text=drawColor,anchor=base,inner sep=0pt, outer sep=0pt, scale=  0.88] at (123.39, 19.68) {2005};

\node[text=drawColor,anchor=base,inner sep=0pt, outer sep=0pt, scale=  0.88] at (190.53, 19.68) {2010};
\end{scope}
\begin{scope}
\path[clip] (  0.00,  0.00) rectangle (245.72,216.81);
\definecolor{drawColor}{RGB}{0,0,0}

\node[text=drawColor,anchor=base,inner sep=0pt, outer sep=0pt, scale=  1.10] at (136.82,  7.70) {crawl year};
\end{scope}
\begin{scope}
\path[clip] (  0.00,  0.00) rectangle (245.72,216.81);
\definecolor{drawColor}{RGB}{0,0,0}

\node[text=drawColor,rotate= 90.00,anchor=base,inner sep=0pt, outer sep=0pt, scale=  1.10] at ( 15.28,100.97) {\% links in archive};
\end{scope}
\begin{scope}
\path[clip] (  0.00,  0.00) rectangle (245.72,216.81);
\definecolor{fillColor}{RGB}{255,255,255}

\path[fill=fillColor] ( 83.25,179.78) rectangle (190.39,202.77);
\end{scope}
\begin{scope}
\path[clip] (  0.00,  0.00) rectangle (245.72,216.81);
\definecolor{drawColor}{RGB}{255,255,255}
\definecolor{fillColor}{gray}{0.95}

\path[draw=drawColor,line width= 0.6pt,line join=round,line cap=round,fill=fillColor] ( 91.14,184.05) rectangle (105.59,198.51);
\end{scope}
\begin{scope}
\path[clip] (  0.00,  0.00) rectangle (245.72,216.81);
\definecolor{drawColor}{RGB}{248,118,109}

\path[draw=drawColor,line width= 2.3pt,line join=round] ( 92.58,191.28) -- (104.14,191.28);
\end{scope}
\begin{scope}
\path[clip] (  0.00,  0.00) rectangle (245.72,216.81);
\definecolor{drawColor}{RGB}{255,255,255}
\definecolor{fillColor}{gray}{0.95}

\path[draw=drawColor,line width= 0.6pt,line join=round,line cap=round,fill=fillColor] (138.80,184.05) rectangle (153.25,198.51);
\end{scope}
\begin{scope}
\path[clip] (  0.00,  0.00) rectangle (245.72,216.81);
\definecolor{drawColor}{RGB}{0,191,196}

\path[draw=drawColor,line width= 2.3pt,dash pattern=on 2pt off 2pt ,line join=round] (140.24,191.28) -- (151.81,191.28);
\end{scope}
\begin{scope}
\path[clip] (  0.00,  0.00) rectangle (245.72,216.81);
\definecolor{drawColor}{RGB}{0,0,0}

\node[text=drawColor,anchor=base west,inner sep=0pt, outer sep=0pt, scale=  0.88] at (107.40,188.25) {internal};
\end{scope}
\begin{scope}
\path[clip] (  0.00,  0.00) rectangle (245.72,216.81);
\definecolor{drawColor}{RGB}{0,0,0}

\node[text=drawColor,anchor=base west,inner sep=0pt, outer sep=0pt, scale=  0.88] at (155.06,188.25) {external};
\end{scope}
\end{tikzpicture}

%% file: figs/features.tex
\begin{tikzpicture}[x=1pt,y=1pt]
\definecolor{fillColor}{RGB}{255,255,255}
\path[use as bounding box,fill=fillColor,fill opacity=0.00] (0,0) rectangle (245.72,216.81);
\begin{scope}
\path[clip] (  0.00,  0.00) rectangle (245.72,216.81);
\definecolor{drawColor}{RGB}{255,255,255}
\definecolor{fillColor}{RGB}{255,255,255}

\path[draw=drawColor,line width= 0.6pt,line join=round,line cap=round,fill=fillColor] (  0.00,  0.00) rectangle (245.72,216.81);
\end{scope}
\begin{scope}
\path[clip] ( 37.82, 30.69) rectangle (240.22,171.25);
\definecolor{fillColor}{gray}{0.92}

\path[fill=fillColor] ( 37.82, 30.69) rectangle (240.22,171.25);
\definecolor{drawColor}{RGB}{255,255,255}

\path[draw=drawColor,line width= 0.3pt,line join=round] ( 37.82, 53.05) --
	(240.22, 53.05);

\path[draw=drawColor,line width= 0.3pt,line join=round] ( 37.82, 84.99) --
	(240.22, 84.99);

\path[draw=drawColor,line width= 0.3pt,line join=round] ( 37.82,116.94) --
	(240.22,116.94);

\path[draw=drawColor,line width= 0.3pt,line join=round] ( 37.82,148.89) --
	(240.22,148.89);

\path[draw=drawColor,line width= 0.3pt,line join=round] ( 63.26, 30.69) --
	( 63.26,171.25);

\path[draw=drawColor,line width= 0.3pt,line join=round] (117.37, 30.69) --
	(117.37,171.25);

\path[draw=drawColor,line width= 0.3pt,line join=round] (171.49, 30.69) --
	(171.49,171.25);

\path[draw=drawColor,line width= 0.3pt,line join=round] (225.61, 30.69) --
	(225.61,171.25);

\path[draw=drawColor,line width= 0.6pt,line join=round] ( 37.82, 37.08) --
	(240.22, 37.08);

\path[draw=drawColor,line width= 0.6pt,line join=round] ( 37.82, 69.02) --
	(240.22, 69.02);

\path[draw=drawColor,line width= 0.6pt,line join=round] ( 37.82,100.97) --
	(240.22,100.97);

\path[draw=drawColor,line width= 0.6pt,line join=round] ( 37.82,132.91) --
	(240.22,132.91);

\path[draw=drawColor,line width= 0.6pt,line join=round] ( 37.82,164.86) --
	(240.22,164.86);

\path[draw=drawColor,line width= 0.6pt,line join=round] ( 90.32, 30.69) --
	( 90.32,171.25);

\path[draw=drawColor,line width= 0.6pt,line join=round] (144.43, 30.69) --
	(144.43,171.25);

\path[draw=drawColor,line width= 0.6pt,line join=round] (198.55, 30.69) --
	(198.55,171.25);
\definecolor{drawColor}{RGB}{248,118,109}

\path[draw=drawColor,line width= 2.3pt,line join=round] ( 47.02, 37.08) --
	( 57.85, 37.43) --
	( 68.67, 37.77) --
	( 79.49, 39.89) --
	( 90.32, 40.93) --
	(101.14, 46.65) --
	(111.96, 51.49) --
	(122.79, 57.77) --
	(133.61, 57.66) --
	(144.43, 66.34) --
	(155.26, 70.99) --
	(166.08, 85.32) --
	(176.90,117.60) --
	(187.73,131.27) --
	(198.55,139.02) --
	(209.37,153.33) --
	(220.19,164.26) --
	(231.02,164.86);
\definecolor{drawColor}{RGB}{163,165,0}

\path[draw=drawColor,line width= 2.3pt,dash pattern=on 2pt off 2pt ,line join=round] ( 47.02, 37.08) --
	( 57.85, 37.08) --
	( 68.67, 37.08) --
	( 79.49, 41.04) --
	( 90.32, 47.96) --
	(101.14, 65.85) --
	(111.96, 75.70) --
	(122.79, 82.75) --
	(133.61, 83.56) --
	(144.43, 92.87) --
	(155.26,104.94) --
	(166.08,136.20) --
	(176.90,164.86) --
	(187.73,144.66) --
	(198.55,150.46) --
	(209.37,153.82) --
	(220.19,155.71) --
	(231.02,136.94);
\definecolor{drawColor}{RGB}{0,191,125}

\path[draw=drawColor,line width= 2.3pt,dash pattern=on 4pt off 2pt ,line join=round] ( 47.02, 37.08) --
	( 57.85, 37.99) --
	( 68.67, 39.42) --
	( 79.49, 40.84) --
	( 90.32, 40.95) --
	(101.14, 46.46) --
	(111.96, 53.14) --
	(122.79, 57.39) --
	(133.61, 60.18) --
	(144.43, 69.30) --
	(155.26, 76.70) --
	(166.08, 99.84) --
	(176.90,113.55) --
	(187.73,135.32) --
	(198.55,152.92) --
	(209.37,158.75) --
	(220.19,164.86) --
	(231.02,161.83);
\definecolor{drawColor}{RGB}{0,176,246}

\path[draw=drawColor,line width= 2.3pt,dash pattern=on 4pt off 4pt ,line join=round] ( 47.02, 37.08) --
	( 57.85, 42.73) --
	( 68.67, 55.34) --
	( 79.49, 65.81) --
	( 90.32, 84.17) --
	(101.14, 90.37) --
	(111.96,129.06) --
	(122.79,122.73) --
	(133.61,115.05) --
	(144.43,145.22) --
	(155.26,154.20) --
	(166.08,145.08) --
	(176.90,164.86) --
	(187.73,133.68) --
	(198.55,121.26) --
	(209.37,115.18) --
	(220.19,103.63) --
	(231.02,100.26);
\definecolor{drawColor}{RGB}{231,107,243}

\path[draw=drawColor,line width= 2.3pt,dash pattern=on 1pt off 3pt ,line join=round] ( 47.02, 37.08) --
	( 57.85, 37.08) --
	( 68.67, 37.21) --
	( 79.49, 38.41) --
	( 90.32, 38.16) --
	(101.14, 40.26) --
	(111.96, 41.82) --
	(122.79, 43.21) --
	(133.61, 46.30) --
	(144.43, 83.29) --
	(155.26, 64.63) --
	(166.08, 73.52) --
	(176.90, 95.12) --
	(187.73,120.95) --
	(198.55,124.07) --
	(209.37,147.99) --
	(220.19,162.59) --
	(231.02,164.86);
\end{scope}
\begin{scope}
\path[clip] (  0.00,  0.00) rectangle (245.72,216.81);
\definecolor{drawColor}{gray}{0.30}

\node[text=drawColor,anchor=base east,inner sep=0pt, outer sep=0pt, scale=  0.88] at ( 32.87, 34.05) {0};

\node[text=drawColor,anchor=base east,inner sep=0pt, outer sep=0pt, scale=  0.88] at ( 32.87, 65.99) {25};

\node[text=drawColor,anchor=base east,inner sep=0pt, outer sep=0pt, scale=  0.88] at ( 32.87, 97.94) {50};

\node[text=drawColor,anchor=base east,inner sep=0pt, outer sep=0pt, scale=  0.88] at ( 32.87,129.88) {75};

\node[text=drawColor,anchor=base east,inner sep=0pt, outer sep=0pt, scale=  0.88] at ( 32.87,161.83) {100};
\end{scope}
\begin{scope}
\path[clip] (  0.00,  0.00) rectangle (245.72,216.81);
\definecolor{drawColor}{gray}{0.20}

\path[draw=drawColor,line width= 0.6pt,line join=round] ( 35.07, 37.08) --
	( 37.82, 37.08);

\path[draw=drawColor,line width= 0.6pt,line join=round] ( 35.07, 69.02) --
	( 37.82, 69.02);

\path[draw=drawColor,line width= 0.6pt,line join=round] ( 35.07,100.97) --
	( 37.82,100.97);

\path[draw=drawColor,line width= 0.6pt,line join=round] ( 35.07,132.91) --
	( 37.82,132.91);

\path[draw=drawColor,line width= 0.6pt,line join=round] ( 35.07,164.86) --
	( 37.82,164.86);
\end{scope}
\begin{scope}
\path[clip] (  0.00,  0.00) rectangle (245.72,216.81);
\definecolor{drawColor}{gray}{0.20}

\path[draw=drawColor,line width= 0.6pt,line join=round] ( 90.32, 27.94) --
	( 90.32, 30.69);

\path[draw=drawColor,line width= 0.6pt,line join=round] (144.43, 27.94) --
	(144.43, 30.69);

\path[draw=drawColor,line width= 0.6pt,line join=round] (198.55, 27.94) --
	(198.55, 30.69);
\end{scope}
\begin{scope}
\path[clip] (  0.00,  0.00) rectangle (245.72,216.81);
\definecolor{drawColor}{gray}{0.30}

\node[text=drawColor,anchor=base,inner sep=0pt, outer sep=0pt, scale=  0.88] at ( 90.32, 19.68) {2000};

\node[text=drawColor,anchor=base,inner sep=0pt, outer sep=0pt, scale=  0.88] at (144.43, 19.68) {2005};

\node[text=drawColor,anchor=base,inner sep=0pt, outer sep=0pt, scale=  0.88] at (198.55, 19.68) {2010};
\end{scope}
\begin{scope}
\path[clip] (  0.00,  0.00) rectangle (245.72,216.81);
\definecolor{drawColor}{RGB}{0,0,0}

\node[text=drawColor,anchor=base,inner sep=0pt, outer sep=0pt, scale=  1.10] at (139.02,  7.70) {crawl year};
\end{scope}
\begin{scope}
\path[clip] (  0.00,  0.00) rectangle (245.72,216.81);
\definecolor{drawColor}{RGB}{0,0,0}

\node[text=drawColor,rotate= 90.00,anchor=base,inner sep=0pt, outer sep=0pt, scale=  1.10] at ( 15.28,100.97) {prevalence (\% of feature max)};
\end{scope}
\begin{scope}
\path[clip] (  0.00,  0.00) rectangle (245.72,216.81);
\definecolor{fillColor}{RGB}{255,255,255}

\path[fill=fillColor] ( 23.10,179.78) rectangle (254.94,202.77);
\end{scope}
\begin{scope}
\path[clip] (  0.00,  0.00) rectangle (245.72,216.81);
\definecolor{drawColor}{RGB}{255,255,255}
\definecolor{fillColor}{gray}{0.95}

\path[draw=drawColor,line width= 0.6pt,line join=round,line cap=round,fill=fillColor] ( 30.98,184.05) rectangle ( 45.43,198.51);
\end{scope}
\begin{scope}
\path[clip] (  0.00,  0.00) rectangle (245.72,216.81);
\definecolor{drawColor}{RGB}{248,118,109}

\path[draw=drawColor,line width= 2.3pt,line join=round] ( 32.43,191.28) -- ( 43.99,191.28);
\end{scope}
\begin{scope}
\path[clip] (  0.00,  0.00) rectangle (245.72,216.81);
\definecolor{drawColor}{RGB}{255,255,255}
\definecolor{fillColor}{gray}{0.95}

\path[draw=drawColor,line width= 0.6pt,line join=round,line cap=round,fill=fillColor] ( 74.10,184.05) rectangle ( 88.55,198.51);
\end{scope}
\begin{scope}
\path[clip] (  0.00,  0.00) rectangle (245.72,216.81);
\definecolor{drawColor}{RGB}{163,165,0}

\path[draw=drawColor,line width= 2.3pt,dash pattern=on 2pt off 2pt ,line join=round] ( 75.54,191.28) -- ( 87.11,191.28);
\end{scope}
\begin{scope}
\path[clip] (  0.00,  0.00) rectangle (245.72,216.81);
\definecolor{drawColor}{RGB}{255,255,255}
\definecolor{fillColor}{gray}{0.95}

\path[draw=drawColor,line width= 0.6pt,line join=round,line cap=round,fill=fillColor] (113.28,184.05) rectangle (127.73,198.51);
\end{scope}
\begin{scope}
\path[clip] (  0.00,  0.00) rectangle (245.72,216.81);
\definecolor{drawColor}{RGB}{0,191,125}

\path[draw=drawColor,line width= 2.3pt,dash pattern=on 4pt off 2pt ,line join=round] (114.73,191.28) -- (126.29,191.28);
\end{scope}
\begin{scope}
\path[clip] (  0.00,  0.00) rectangle (245.72,216.81);
\definecolor{drawColor}{RGB}{255,255,255}
\definecolor{fillColor}{gray}{0.95}

\path[draw=drawColor,line width= 0.6pt,line join=round,line cap=round,fill=fillColor] (178.37,184.05) rectangle (192.82,198.51);
\end{scope}
\begin{scope}
\path[clip] (  0.00,  0.00) rectangle (245.72,216.81);
\definecolor{drawColor}{RGB}{0,176,246}

\path[draw=drawColor,line width= 2.3pt,dash pattern=on 4pt off 4pt ,line join=round] (179.81,191.28) -- (191.38,191.28);
\end{scope}
\begin{scope}
\path[clip] (  0.00,  0.00) rectangle (245.72,216.81);
\definecolor{drawColor}{RGB}{255,255,255}
\definecolor{fillColor}{gray}{0.95}

\path[draw=drawColor,line width= 0.6pt,line join=round,line cap=round,fill=fillColor] (218.97,184.05) rectangle (233.42,198.51);
\end{scope}
\begin{scope}
\path[clip] (  0.00,  0.00) rectangle (245.72,216.81);
\definecolor{drawColor}{RGB}{231,107,243}

\path[draw=drawColor,line width= 2.3pt,dash pattern=on 1pt off 3pt ,line join=round] (220.41,191.28) -- (231.98,191.28);
\end{scope}
\begin{scope}
\path[clip] (  0.00,  0.00) rectangle (245.72,216.81);
\definecolor{drawColor}{RGB}{0,0,0}

\node[text=drawColor,anchor=base west,inner sep=0pt, outer sep=0pt, scale=  0.88] at ( 47.24,188.25) {scripts};
\end{scope}
\begin{scope}
\path[clip] (  0.00,  0.00) rectangle (245.72,216.81);
\definecolor{drawColor}{RGB}{0,0,0}

\node[text=drawColor,anchor=base west,inner sep=0pt, outer sep=0pt, scale=  0.88] at ( 90.36,188.25) {styles};
\end{scope}
\begin{scope}
\path[clip] (  0.00,  0.00) rectangle (245.72,216.81);
\definecolor{drawColor}{RGB}{0,0,0}

\node[text=drawColor,anchor=base west,inner sep=0pt, outer sep=0pt, scale=  0.88] at (129.54,188.25) {linked styles};
\end{scope}
\begin{scope}
\path[clip] (  0.00,  0.00) rectangle (245.72,216.81);
\definecolor{drawColor}{RGB}{0,0,0}

\node[text=drawColor,anchor=base west,inner sep=0pt, outer sep=0pt, scale=  0.88] at (194.63,188.25) {tables};
\end{scope}
\begin{scope}
\path[clip] (  0.00,  0.00) rectangle (245.72,216.81);
\definecolor{drawColor}{RGB}{0,0,0}

\node[text=drawColor,anchor=base west,inner sep=0pt, outer sep=0pt, scale=  0.88] at (235.23,188.25) {divs};
\end{scope}
\end{tikzpicture}

%% file: figs/features_t.tex
\begin{tikzpicture}[x=1pt,y=1pt]
\definecolor{fillColor}{RGB}{255,255,255}
\path[use as bounding box,fill=fillColor,fill opacity=0.00] (0,0) rectangle (245.72,216.81);
\begin{scope}
\path[clip] (  0.00,  0.00) rectangle (245.72,216.81);
\definecolor{drawColor}{RGB}{255,255,255}
\definecolor{fillColor}{RGB}{255,255,255}

\path[draw=drawColor,line width= 0.6pt,line join=round,line cap=round,fill=fillColor] (  0.00,  0.00) rectangle (245.72,216.81);
\end{scope}
\begin{scope}
\path[clip] ( 33.42, 30.69) rectangle (240.22,171.25);
\definecolor{fillColor}{gray}{0.92}

\path[fill=fillColor] ( 33.42, 30.69) rectangle (240.22,171.25);
\definecolor{drawColor}{RGB}{255,255,255}

\path[draw=drawColor,line width= 0.3pt,line join=round] ( 33.42, 61.30) --
	(240.22, 61.30);

\path[draw=drawColor,line width= 0.3pt,line join=round] ( 33.42, 97.63) --
	(240.22, 97.63);

\path[draw=drawColor,line width= 0.3pt,line join=round] ( 33.42,137.99) --
	(240.22,137.99);

\path[draw=drawColor,line width= 0.3pt,line join=round] ( 59.41, 30.69) --
	( 59.41,171.25);

\path[draw=drawColor,line width= 0.3pt,line join=round] (114.70, 30.69) --
	(114.70,171.25);

\path[draw=drawColor,line width= 0.3pt,line join=round] (170.00, 30.69) --
	(170.00,171.25);

\path[draw=drawColor,line width= 0.3pt,line join=round] (225.29, 30.69) --
	(225.29,171.25);

\path[draw=drawColor,line width= 0.6pt,line join=round] ( 33.42, 45.15) --
	(240.22, 45.15);

\path[draw=drawColor,line width= 0.6pt,line join=round] ( 33.42, 77.44) --
	(240.22, 77.44);

\path[draw=drawColor,line width= 0.6pt,line join=round] ( 33.42,117.81) --
	(240.22,117.81);

\path[draw=drawColor,line width= 0.6pt,line join=round] ( 33.42,158.18) --
	(240.22,158.18);

\path[draw=drawColor,line width= 0.6pt,line join=round] ( 87.06, 30.69) --
	( 87.06,171.25);

\path[draw=drawColor,line width= 0.6pt,line join=round] (142.35, 30.69) --
	(142.35,171.25);

\path[draw=drawColor,line width= 0.6pt,line join=round] (197.64, 30.69) --
	(197.64,171.25);
\definecolor{drawColor}{RGB}{248,118,109}

\path[draw=drawColor,line width= 2.3pt,line join=round] ( 42.82, 37.08) --
	( 53.88, 37.43) --
	( 64.94, 37.77) --
	( 76.00, 39.89) --
	( 87.06, 40.93) --
	( 98.12, 46.65) --
	(109.17, 51.49) --
	(120.23, 57.77) --
	(131.29, 57.66) --
	(142.35, 66.34) --
	(153.41, 70.99) --
	(164.47, 85.32) --
	(175.53,117.60) --
	(186.58,131.27) --
	(197.64,139.02) --
	(208.70,153.33) --
	(219.76,164.26) --
	(230.82,164.86);
\definecolor{drawColor}{RGB}{0,186,56}

\path[draw=drawColor,line width= 2.3pt,dash pattern=on 2pt off 2pt ,line join=round] ( 42.82, 37.08) --
	( 53.88, 37.08) --
	( 64.94, 37.08) --
	( 76.00, 37.30) --
	( 87.06, 37.69) --
	( 98.12, 38.69) --
	(109.17, 39.24) --
	(120.23, 39.63) --
	(131.29, 39.68) --
	(142.35, 40.20) --
	(153.41, 40.87) --
	(164.47, 42.63) --
	(175.53, 44.23) --
	(186.58, 43.10) --
	(197.64, 43.42) --
	(208.70, 43.61) --
	(219.76, 43.72) --
	(230.82, 42.67);
\definecolor{drawColor}{RGB}{97,156,255}

\path[draw=drawColor,line width= 2.3pt,dash pattern=on 4pt off 2pt ,line join=round] ( 42.82, 37.08) --
	( 53.88, 37.26) --
	( 64.94, 37.54) --
	( 76.00, 37.82) --
	( 87.06, 37.84) --
	( 98.12, 38.92) --
	(109.17, 40.23) --
	(120.23, 41.07) --
	(131.29, 41.62) --
	(142.35, 43.41) --
	(153.41, 44.86) --
	(164.47, 49.41) --
	(175.53, 52.11) --
	(186.58, 56.39) --
	(197.64, 59.85) --
	(208.70, 60.99) --
	(219.76, 62.19) --
	(230.82, 61.60);
\end{scope}
\begin{scope}
\path[clip] (  0.00,  0.00) rectangle (245.72,216.81);
\definecolor{drawColor}{gray}{0.30}

\node[text=drawColor,anchor=base east,inner sep=0pt, outer sep=0pt, scale=  0.88] at ( 28.47, 42.12) {1};

\node[text=drawColor,anchor=base east,inner sep=0pt, outer sep=0pt, scale=  0.88] at ( 28.47, 74.41) {5};

\node[text=drawColor,anchor=base east,inner sep=0pt, outer sep=0pt, scale=  0.88] at ( 28.47,114.78) {10};

\node[text=drawColor,anchor=base east,inner sep=0pt, outer sep=0pt, scale=  0.88] at ( 28.47,155.15) {15};
\end{scope}
\begin{scope}
\path[clip] (  0.00,  0.00) rectangle (245.72,216.81);
\definecolor{drawColor}{gray}{0.20}

\path[draw=drawColor,line width= 0.6pt,line join=round] ( 30.67, 45.15) --
	( 33.42, 45.15);

\path[draw=drawColor,line width= 0.6pt,line join=round] ( 30.67, 77.44) --
	( 33.42, 77.44);

\path[draw=drawColor,line width= 0.6pt,line join=round] ( 30.67,117.81) --
	( 33.42,117.81);

\path[draw=drawColor,line width= 0.6pt,line join=round] ( 30.67,158.18) --
	( 33.42,158.18);
\end{scope}
\begin{scope}
\path[clip] (  0.00,  0.00) rectangle (245.72,216.81);
\definecolor{drawColor}{gray}{0.20}

\path[draw=drawColor,line width= 0.6pt,line join=round] ( 87.06, 27.94) --
	( 87.06, 30.69);

\path[draw=drawColor,line width= 0.6pt,line join=round] (142.35, 27.94) --
	(142.35, 30.69);

\path[draw=drawColor,line width= 0.6pt,line join=round] (197.64, 27.94) --
	(197.64, 30.69);
\end{scope}
\begin{scope}
\path[clip] (  0.00,  0.00) rectangle (245.72,216.81);
\definecolor{drawColor}{gray}{0.30}

\node[text=drawColor,anchor=base,inner sep=0pt, outer sep=0pt, scale=  0.88] at ( 87.06, 19.68) {2000};

\node[text=drawColor,anchor=base,inner sep=0pt, outer sep=0pt, scale=  0.88] at (142.35, 19.68) {2005};

\node[text=drawColor,anchor=base,inner sep=0pt, outer sep=0pt, scale=  0.88] at (197.64, 19.68) {2010};
\end{scope}
\begin{scope}
\path[clip] (  0.00,  0.00) rectangle (245.72,216.81);
\definecolor{drawColor}{RGB}{0,0,0}

\node[text=drawColor,anchor=base,inner sep=0pt, outer sep=0pt, scale=  1.10] at (136.82,  7.70) {crawl year};
\end{scope}
\begin{scope}
\path[clip] (  0.00,  0.00) rectangle (245.72,216.81);
\definecolor{drawColor}{RGB}{0,0,0}

\node[text=drawColor,rotate= 90.00,anchor=base,inner sep=0pt, outer sep=0pt, scale=  1.10] at ( 15.28,100.97) {avg. number of tags per page};
\end{scope}
\begin{scope}
\path[clip] (  0.00,  0.00) rectangle (245.72,216.81);
\definecolor{fillColor}{RGB}{255,255,255}

\path[fill=fillColor] ( 57.96,179.78) rectangle (215.69,202.77);
\end{scope}
\begin{scope}
\path[clip] (  0.00,  0.00) rectangle (245.72,216.81);
\definecolor{drawColor}{RGB}{255,255,255}
\definecolor{fillColor}{gray}{0.95}

\path[draw=drawColor,line width= 0.6pt,line join=round,line cap=round,fill=fillColor] ( 65.84,184.05) rectangle ( 80.29,198.51);
\end{scope}
\begin{scope}
\path[clip] (  0.00,  0.00) rectangle (245.72,216.81);
\definecolor{drawColor}{RGB}{248,118,109}

\path[draw=drawColor,line width= 2.3pt,line join=round] ( 67.28,191.28) -- ( 78.85,191.28);
\end{scope}
\begin{scope}
\path[clip] (  0.00,  0.00) rectangle (245.72,216.81);
\definecolor{drawColor}{RGB}{255,255,255}
\definecolor{fillColor}{gray}{0.95}

\path[draw=drawColor,line width= 0.6pt,line join=round,line cap=round,fill=fillColor] (108.95,184.05) rectangle (123.41,198.51);
\end{scope}
\begin{scope}
\path[clip] (  0.00,  0.00) rectangle (245.72,216.81);
\definecolor{drawColor}{RGB}{0,186,56}

\path[draw=drawColor,line width= 2.3pt,dash pattern=on 2pt off 2pt ,line join=round] (110.40,191.28) -- (121.96,191.28);
\end{scope}
\begin{scope}
\path[clip] (  0.00,  0.00) rectangle (245.72,216.81);
\definecolor{drawColor}{RGB}{255,255,255}
\definecolor{fillColor}{gray}{0.95}

\path[draw=drawColor,line width= 0.6pt,line join=round,line cap=round,fill=fillColor] (148.14,184.05) rectangle (162.59,198.51);
\end{scope}
\begin{scope}
\path[clip] (  0.00,  0.00) rectangle (245.72,216.81);
\definecolor{drawColor}{RGB}{97,156,255}

\path[draw=drawColor,line width= 2.3pt,dash pattern=on 4pt off 2pt ,line join=round] (149.58,191.28) -- (161.15,191.28);
\end{scope}
\begin{scope}
\path[clip] (  0.00,  0.00) rectangle (245.72,216.81);
\definecolor{drawColor}{RGB}{0,0,0}

\node[text=drawColor,anchor=base west,inner sep=0pt, outer sep=0pt, scale=  0.88] at ( 82.10,188.25) {scripts};
\end{scope}
\begin{scope}
\path[clip] (  0.00,  0.00) rectangle (245.72,216.81);
\definecolor{drawColor}{RGB}{0,0,0}

\node[text=drawColor,anchor=base west,inner sep=0pt, outer sep=0pt, scale=  0.88] at (125.22,188.25) {styles};
\end{scope}
\begin{scope}
\path[clip] (  0.00,  0.00) rectangle (245.72,216.81);
\definecolor{drawColor}{RGB}{0,0,0}

\node[text=drawColor,anchor=base west,inner sep=0pt, outer sep=0pt, scale=  0.88] at (164.40,188.25) {linked styles};
\end{scope}
\end{tikzpicture}

%% file: figs/links.tex
\begin{tikzpicture}[x=1pt,y=1pt]
\definecolor{fillColor}{RGB}{255,255,255}
\path[use as bounding box,fill=fillColor,fill opacity=0.00] (0,0) rectangle (245.72,216.81);
\begin{scope}
\path[clip] (  0.00,  0.00) rectangle (245.72,216.81);
\definecolor{drawColor}{RGB}{255,255,255}
\definecolor{fillColor}{RGB}{255,255,255}

\path[draw=drawColor,line width= 0.6pt,line join=round,line cap=round,fill=fillColor] (  0.00,  0.00) rectangle (245.72,216.81);
\end{scope}
\begin{scope}
\path[clip] ( 37.82, 30.69) rectangle (240.22,171.25);
\definecolor{fillColor}{gray}{0.92}

\path[fill=fillColor] ( 37.82, 30.69) rectangle (240.22,171.25);
\definecolor{drawColor}{RGB}{255,255,255}

\path[draw=drawColor,line width= 0.3pt,line join=round] ( 37.82, 48.03) --
	(240.22, 48.03);

\path[draw=drawColor,line width= 0.3pt,line join=round] ( 37.82, 74.95) --
	(240.22, 74.95);

\path[draw=drawColor,line width= 0.3pt,line join=round] ( 37.82,101.87) --
	(240.22,101.87);

\path[draw=drawColor,line width= 0.3pt,line join=round] ( 37.82,128.80) --
	(240.22,128.80);

\path[draw=drawColor,line width= 0.3pt,line join=round] ( 37.82,155.72) --
	(240.22,155.72);

\path[draw=drawColor,line width= 0.3pt,line join=round] ( 93.02, 30.69) --
	( 93.02,171.25);

\path[draw=drawColor,line width= 0.3pt,line join=round] (158.73, 30.69) --
	(158.73,171.25);

\path[draw=drawColor,line width= 0.3pt,line join=round] (224.45, 30.69) --
	(224.45,171.25);

\path[draw=drawColor,line width= 0.6pt,line join=round] ( 37.82, 34.56) --
	(240.22, 34.56);

\path[draw=drawColor,line width= 0.6pt,line join=round] ( 37.82, 61.49) --
	(240.22, 61.49);

\path[draw=drawColor,line width= 0.6pt,line join=round] ( 37.82, 88.41) --
	(240.22, 88.41);

\path[draw=drawColor,line width= 0.6pt,line join=round] ( 37.82,115.34) --
	(240.22,115.34);

\path[draw=drawColor,line width= 0.6pt,line join=round] ( 37.82,142.26) --
	(240.22,142.26);

\path[draw=drawColor,line width= 0.6pt,line join=round] ( 37.82,169.18) --
	(240.22,169.18);

\path[draw=drawColor,line width= 0.6pt,line join=round] ( 60.17, 30.69) --
	( 60.17,171.25);

\path[draw=drawColor,line width= 0.6pt,line join=round] (125.88, 30.69) --
	(125.88,171.25);

\path[draw=drawColor,line width= 0.6pt,line join=round] (191.59, 30.69) --
	(191.59,171.25);
\definecolor{drawColor}{RGB}{248,118,109}

\path[draw=drawColor,line width= 2.3pt,line join=round] ( 47.02, 48.66) --
	( 60.17, 71.01) --
	( 73.31, 60.27) --
	( 86.45, 63.31) --
	( 99.59, 69.98) --
	(112.74, 84.17) --
	(125.88, 98.68) --
	(139.02, 96.51) --
	(152.16, 85.31) --
	(165.31,110.55) --
	(178.45,115.30) --
	(191.59,122.69) --
	(204.73,134.62) --
	(217.88,148.97) --
	(231.02,164.86);
\definecolor{drawColor}{RGB}{0,186,56}

\path[draw=drawColor,line width= 2.3pt,dash pattern=on 4pt off 4pt ,line join=round] ( 47.02, 45.68) --
	( 60.17, 68.50) --
	( 73.31, 56.95) --
	( 86.45, 57.96) --
	( 99.59, 63.40) --
	(112.74, 75.02) --
	(125.88, 91.35) --
	(139.02, 88.12) --
	(152.16, 79.28) --
	(165.31,101.18) --
	(178.45,107.25) --
	(191.59,114.03) --
	(204.73,124.53) --
	(217.88,138.39) --
	(231.02,152.05);
\definecolor{drawColor}{RGB}{97,156,255}

\path[draw=drawColor,line width= 2.3pt,dash pattern=on 1pt off 3pt on 4pt off 3pt ,line join=round] ( 47.02, 37.55) --
	( 60.17, 37.08) --
	( 73.31, 37.89) --
	( 86.45, 39.92) --
	( 99.59, 41.15) --
	(112.74, 43.71) --
	(125.88, 41.90) --
	(139.02, 42.95) --
	(152.16, 40.59) --
	(165.31, 43.93) --
	(178.45, 42.62) --
	(191.59, 43.23) --
	(204.73, 44.65) --
	(217.88, 45.14) --
	(231.02, 47.37);
\end{scope}
\begin{scope}
\path[clip] (  0.00,  0.00) rectangle (245.72,216.81);
\definecolor{drawColor}{gray}{0.30}

\node[text=drawColor,anchor=base east,inner sep=0pt, outer sep=0pt, scale=  0.88] at ( 32.87, 31.53) {0};

\node[text=drawColor,anchor=base east,inner sep=0pt, outer sep=0pt, scale=  0.88] at ( 32.87, 58.46) {25};

\node[text=drawColor,anchor=base east,inner sep=0pt, outer sep=0pt, scale=  0.88] at ( 32.87, 85.38) {50};

\node[text=drawColor,anchor=base east,inner sep=0pt, outer sep=0pt, scale=  0.88] at ( 32.87,112.31) {75};

\node[text=drawColor,anchor=base east,inner sep=0pt, outer sep=0pt, scale=  0.88] at ( 32.87,139.23) {100};

\node[text=drawColor,anchor=base east,inner sep=0pt, outer sep=0pt, scale=  0.88] at ( 32.87,166.15) {125};
\end{scope}
\begin{scope}
\path[clip] (  0.00,  0.00) rectangle (245.72,216.81);
\definecolor{drawColor}{gray}{0.20}

\path[draw=drawColor,line width= 0.6pt,line join=round] ( 35.07, 34.56) --
	( 37.82, 34.56);

\path[draw=drawColor,line width= 0.6pt,line join=round] ( 35.07, 61.49) --
	( 37.82, 61.49);

\path[draw=drawColor,line width= 0.6pt,line join=round] ( 35.07, 88.41) --
	( 37.82, 88.41);

\path[draw=drawColor,line width= 0.6pt,line join=round] ( 35.07,115.34) --
	( 37.82,115.34);

\path[draw=drawColor,line width= 0.6pt,line join=round] ( 35.07,142.26) --
	( 37.82,142.26);

\path[draw=drawColor,line width= 0.6pt,line join=round] ( 35.07,169.18) --
	( 37.82,169.18);
\end{scope}
\begin{scope}
\path[clip] (  0.00,  0.00) rectangle (245.72,216.81);
\definecolor{drawColor}{gray}{0.20}

\path[draw=drawColor,line width= 0.6pt,line join=round] ( 60.17, 27.94) --
	( 60.17, 30.69);

\path[draw=drawColor,line width= 0.6pt,line join=round] (125.88, 27.94) --
	(125.88, 30.69);

\path[draw=drawColor,line width= 0.6pt,line join=round] (191.59, 27.94) --
	(191.59, 30.69);
\end{scope}
\begin{scope}
\path[clip] (  0.00,  0.00) rectangle (245.72,216.81);
\definecolor{drawColor}{gray}{0.30}

\node[text=drawColor,anchor=base,inner sep=0pt, outer sep=0pt, scale=  0.88] at ( 60.17, 19.68) {2000};

\node[text=drawColor,anchor=base,inner sep=0pt, outer sep=0pt, scale=  0.88] at (125.88, 19.68) {2005};

\node[text=drawColor,anchor=base,inner sep=0pt, outer sep=0pt, scale=  0.88] at (191.59, 19.68) {2010};
\end{scope}
\begin{scope}
\path[clip] (  0.00,  0.00) rectangle (245.72,216.81);
\definecolor{drawColor}{RGB}{0,0,0}

\node[text=drawColor,anchor=base,inner sep=0pt, outer sep=0pt, scale=  1.10] at (139.02,  7.70) {crawl year};
\end{scope}
\begin{scope}
\path[clip] (  0.00,  0.00) rectangle (245.72,216.81);
\definecolor{drawColor}{RGB}{0,0,0}

\node[text=drawColor,rotate= 90.00,anchor=base,inner sep=0pt, outer sep=0pt, scale=  1.10] at ( 15.28,100.97) {average links per document};
\end{scope}
\begin{scope}
\path[clip] (  0.00,  0.00) rectangle (245.72,216.81);
\definecolor{fillColor}{RGB}{255,255,255}

\path[fill=fillColor] ( 67.38,179.78) rectangle (210.66,202.77);
\end{scope}
\begin{scope}
\path[clip] (  0.00,  0.00) rectangle (245.72,216.81);
\definecolor{drawColor}{RGB}{255,255,255}
\definecolor{fillColor}{gray}{0.95}

\path[draw=drawColor,line width= 0.6pt,line join=round,line cap=round,fill=fillColor] ( 75.26,184.05) rectangle ( 89.71,198.51);
\end{scope}
\begin{scope}
\path[clip] (  0.00,  0.00) rectangle (245.72,216.81);
\definecolor{drawColor}{RGB}{248,118,109}

\path[draw=drawColor,line width= 2.3pt,line join=round] ( 76.70,191.28) -- ( 88.27,191.28);
\end{scope}
\begin{scope}
\path[clip] (  0.00,  0.00) rectangle (245.72,216.81);
\definecolor{drawColor}{RGB}{255,255,255}
\definecolor{fillColor}{gray}{0.95}

\path[draw=drawColor,line width= 0.6pt,line join=round,line cap=round,fill=fillColor] (111.41,184.05) rectangle (125.86,198.51);
\end{scope}
\begin{scope}
\path[clip] (  0.00,  0.00) rectangle (245.72,216.81);
\definecolor{drawColor}{RGB}{0,186,56}

\path[draw=drawColor,line width= 2.3pt,dash pattern=on 4pt off 4pt ,line join=round] (112.86,191.28) -- (124.42,191.28);
\end{scope}
\begin{scope}
\path[clip] (  0.00,  0.00) rectangle (245.72,216.81);
\definecolor{drawColor}{RGB}{255,255,255}
\definecolor{fillColor}{gray}{0.95}

\path[draw=drawColor,line width= 0.6pt,line join=round,line cap=round,fill=fillColor] (159.07,184.05) rectangle (173.53,198.51);
\end{scope}
\begin{scope}
\path[clip] (  0.00,  0.00) rectangle (245.72,216.81);
\definecolor{drawColor}{RGB}{97,156,255}

\path[draw=drawColor,line width= 2.3pt,dash pattern=on 1pt off 3pt on 4pt off 3pt ,line join=round] (160.52,191.28) -- (172.08,191.28);
\end{scope}
\begin{scope}
\path[clip] (  0.00,  0.00) rectangle (245.72,216.81);
\definecolor{drawColor}{RGB}{0,0,0}

\node[text=drawColor,anchor=base west,inner sep=0pt, outer sep=0pt, scale=  0.88] at ( 91.52,188.25) {total};
\end{scope}
\begin{scope}
\path[clip] (  0.00,  0.00) rectangle (245.72,216.81);
\definecolor{drawColor}{RGB}{0,0,0}

\node[text=drawColor,anchor=base west,inner sep=0pt, outer sep=0pt, scale=  0.88] at (127.67,188.25) {internal};
\end{scope}
\begin{scope}
\path[clip] (  0.00,  0.00) rectangle (245.72,216.81);
\definecolor{drawColor}{RGB}{0,0,0}

\node[text=drawColor,anchor=base west,inner sep=0pt, outer sep=0pt, scale=  0.88] at (175.33,188.25) {external};
\end{scope}
\end{tikzpicture}